\documentclass[prd,twocolumn,groupedaddress,preprintnumbers, nofootinbib,nobibnotes]{revtex4-1}
\catcode`\@=11
\def\lsim{\mathrel{\mathpalette\@versim<}}
\def\gsim{\mathrel{\mathpalette\@versim>}}
\def\@versim#1#2{\vcenter{\offinterlineskip
\ialign{$\m@th#1\hfil##\hfil$\crcr#2\crcr\sim\crcr } }}
\catcode`\@=12

\catcode`@=12
\usepackage[export]{adjustbox}	
\usepackage{float}
\usepackage{caption}
\usepackage{epsfig,bbm}
\usepackage{amsmath}
\usepackage{slashed}
\usepackage{tikz}
\usepackage{bm}
\usepackage{amssymb}
\usepackage{mathtools}
\usepackage[caption=false]{subfig}
\usepackage{graphicx}
\usepackage{hyperref}
\usepackage{cleveref}

\newcommand{\be}{\begin{equation}}
\newcommand{\ee}{\end{equation}}
\newcommand{\bea}{\begin{eqnarray}}
\newcommand{\eea}{\end{eqnarray}}

\begin{document}
\thispagestyle{empty}


\begin{center}
\title{Beyond the Starobinsky model for inflation
}

 \author{Dhong Yeon Cheong}
\affiliation{%
Department of Physics \& IPAP \& Lab for Dark Universe, Yonsei University, Seoul 03722 Korea
}
 \author{Hyun Min Lee}
 \email[]{hminlee@cau.ac.kr}
 \thanks{co-corresponding author}
\affiliation{%
Department of Physics, Chung-Ang University, Seoul 06974, Korea}
\affiliation{CERN, Theory department, 1211 Geneva 23, Switzerland
}
 \author{Seong Chan Park}
 \email[]{sc.park@yonsei.ac.kr}
 \thanks{co-corresponding author}
\affiliation{%
Department of Physics \& IPAP \& Lab for Dark Universe, Yonsei University, Seoul 03722 Korea
}

\preprint{YHEP-COS20-04,  
CAU-THEP-2020-02,  
CERN-TH-2020-022}

\begin{abstract}
We single out the Starobinsky model and its extensions among generic $f(R)$ gravity as attractors at large field values for chaotic inflation. Treating a $R^3$ curvature term as a perturbation of the Starobinsky model, we impose the phenomenological bounds on the additional term satisfying the successful inflationary predictions. We find that the scalar spectral index can vary in both the red or blue tilted direction, depending on the sign of the coefficient of the $R^3$ term, whereas the tensor-to-scalar ratio is less affected in the Planck-compatible region. We also discuss the role of higher order curvature term for stability and the reheating dynamics for the unambiguous prediction for the number of efoldings up to the $R^3$ term.
  
\end{abstract}
\maketitle
\end{center}

\section{Introduction}

Cosmic inflation solves various problems of standard Big Bang cosmology including the horizon problem, homogeneity, structure formation, etc, and it has been tested by the measurements of Cosmic Microwave Background anisotropies with unprecedented precision. Favored vanilla single-field inflations consist a canonical kinetic term, and some with monomial type potentials have now been excluded at more than $2\sigma$ level by the measured scalar spectral index and the bound on the tensor-to-scalar ratio \cite{Akrami:2018odb}.

The Starobinsky inflation model \cite{Starobinsky:1980te} drew new attention from the fact that a successful slow-roll inflation can be obtained with a single parameter beyond the SM, namely the coefficient of the $R^2$ curvature term.  The inflationary predictions of the Starobinsky model are well consistent with the Planck data. Therefore, the discussion has been generalized to a class of Starobinsky-like models with common properties 
during inflation \cite{Kallosh:2013hoa,Kallosh:2013tua,Kallosh:2013yoa,Kehagias:2013mya,Giudice:2014toa}, including the Higgs inflation as a particular case \cite{Bezrukov:2007ep}. 

The unitarity issue is important in defining the validity of the semi-classical treatment of inflationary dynamics. In the case of the original Higgs inflation with a large non-minimal coupling, the unitarity problem occurs due to the would-be Goldstone components of the Higgs field  \cite{Burgess:2009ea,Barbon:2009ya,Burgess:2010zq,Hertzberg:2010dc}, which motivated sigma-model type extensions \cite{Giudice:2010ka,Barbon:2015fla,Lee:2018esk,Choi:2019osi}. In the case of Higgs inflation at criticality where both the Higgs quartic coupling and its beta function coefficient almost vanish~\cite{Hamada:2014iga, Hamada:2014wna}, the unitarity scale is far above the Hubble scale during inflation, so the unitary problem is much milder. 
 
In the case of the Starobinsky model,  the dynamics of the dual scalar field can unitarize the Higgs inflation up to the Planck scale \cite{Kehagias:2013mya,Giudice:2014toa}. Making an appropriate field redefinition of the dual scalar field and transforming to the Einstein frame, the Starobinsky model provides an appropriate coupling between the dual scalar field and the Higgs field such that Higgs inflation is recovered below the mass of the dual scalar field \cite{Ema:2017rqn,Gorbunov:2018llf,He:2018mgb,Gundhi:2018wyz}.  Other theoretical issues such as fine-tuning~\cite{Park:2018kst}, swampland conjecture~\cite{Cheong:2018udx} and the Palatini formulation of Higgs inflation~\cite{Tenkanen:2019jiq, Jinno:2019und} are also recently addressed.

It has also been shown recently that the nontrivial inflaton trajectory in the Higgs-$R^2$ inflation \cite{Ema:2017rqn,He:2018gyf} can provide an interesting possibility that primordial black holes can form during inflation as the dark matter candidate \cite{Pi:2017gih,Cheong:2019vzl}.
However, in the region of the parameter space where primordial black holes saturate the relic density, the resulting spectral index of the curvature perturbations is slightly more red-tilted as compared to the best-fit value of the Planck data at $1\sigma$ level \cite{Pi:2017gih,Cheong:2019vzl}.

In this article, we discuss the Starobinsky inflation model among general $f(R)$ gravity models from the point of attractors at large fields for chaotic inflation. Extending the Starobinsky model with a cubic $R^3$ curvature term, we impose the conditions on the cubic curvature term for maintaining a successful inflation and identify how the inflationary predictions of the Starobinsky model can be modified. We also briefly discuss the potential instability of the cubic term and the effects of even higher order curvature term on this issue.  The reheating dynamics up to $R^3$ term is also dealt with for completeness.

The article is organized as follows.
We begin with a connection between a generic $f(R)$ gravity and its scalar dual theory. Then, we show the criteria for $f(R)$ gravity to give successful predictions for inflation. Next, we extend the Starobinsky model with a cubic $R^3$ term and derive the inflationary observables as compared to those of the Starobinsky model. We go on to discuss the reheating dynamics up to $R^3$ correction  and show the unambiguous prediction for the number of efoldings in this case. 
We also discuss the roles of the dual scalar field in the extended Starobinsky model for unitarizing the Higgs inflation with a non-minimal coupling and curing the vacuum instability problem in the SM.
Finally, conclusions are drawn.

\section{The dual scalar theory of $f(R)$}

We can connect a generic $f(R)$ gravity to a corresponding scalar-tensor theory by Legendre transformation:
\begin{align}
S&=\frac{1}{2}\int d^4 x \sqrt{-g}\, f(R) \\
\to S&=\frac{1}{2}\int d^4 x \sqrt{-g}\, \left[ f(\phi) + f'(\phi) (R-\phi) \right]\\
&\equiv \int d^4 x \sqrt{-g} \left[ \frac{1}{2}\Omega^2 R -V(\phi)\right]
\end{align}
where the frame function and the potential are respectively given as 
\begin{align}
\Omega^2(\phi) &= f'(\phi),  \label{frame} \\
V(\phi) &= \frac{1}{2}\left[ \phi f'(\phi) -f(\phi) \right].
\end{align}
One notes that the variation $\delta \phi$ of the second equation recovers the original action.

The action in the Einstein frame can be obtained by Weyl transformation 
$g_{E\mu\nu}= \Omega^2 g_{\mu\nu}$:
\begin{align}
S_E =\int d^4 x \sqrt{-g_E} \left[ \frac{1}{2}R_E - \frac{1}{2}g^{\mu\nu}_E \partial_\mu s \partial_\nu s -V_E(s)\right],
\end{align}
where the canonical field $s$ and the potential in the Einstein frame is $V_E$ are respectively given as~\footnote{$\bar{g}_{ab} =e^{2\psi} g_{ab}$ gives $\sqrt{-\bar{g}} =e^{D\psi}\sqrt{-g}$ and $\bar{R} =e^{-2\psi}\left[R-2(D-1)\nabla^2 \psi -(D-2)(D-1)g^{ab}\partial_a \psi \partial_b \psi\right]$ in $D$-dimensions.} 
\begin{align}
s(\phi)&=\sqrt{\frac{3}{2}} \ln \Omega^2(\phi) = \sqrt{\frac{3}{2}} \ln \left[f'(\phi)\right], \\
V_E(s)&= \frac{V(\phi(s))}{\Omega(\phi(s))^4} =\left.\frac{\phi f'(\phi) -f(\phi)}{2 f'^2(\phi)}\right|_{\phi=\phi(s)},
\label{Einsteinpot}
\end{align}
where $\phi(s)$ can be obtained by inverting $s(\phi)$. 

We note that the chaotic inflation constrains the asymptotic form of $f(\phi)$:
for instance, a monomial function $f(\phi)\sim \phi^{n+1}$ leads to 
\begin{align}
V_E \sim \frac{\phi^{n+1}}{\phi^{2n}} \sim \phi^{1-n} 
\end{align}
such that $n=1$ gives a flat potential for inflation.

\section{Selection rules for inflation}

We consider a general form of the higher curvature correction to Einstein gravity by taking $f(R)=a R+ b R^{n+1}$ with $n\geq 1$, and discuss the selection rules for a successful slow-roll inflation.
 
Putting $a=1$ and $b\equiv \beta/(n+1)$, $f'(\phi) = 1+\beta R^n$,
\begin{eqnarray}
s(\phi) &=& \sqrt{\frac{3}{2}}\ln \left[1+ \beta \phi^n\right], \\
\sigma(s) &\equiv&e^{\sqrt{\frac{2}{3}}s}=1+\beta \phi^n.
\end{eqnarray}
The equation is easily solved and we obtain $\phi(s)$:
\begin{align}
\phi(s) &=\left(\frac{\sigma(s)-1}{\beta}\right)^{\tfrac{1}{n}}.
\end{align}
The potential in Einstein frame is
\begin{align}
V_E(s) 
&= \frac{\phi(s)\sigma(s)- f(\phi(s))}{2\sigma(s)^2} \\
&=\frac{n }{2(n+1)\beta^{1/n}} \frac{(\sigma-1)^{\tfrac{n+1}{n}}}{\sigma^2},
\end{align}
where $f(\phi(s))= \left(\frac{\sigma(s)-1}{\beta}\right)^{\tfrac{1}{n}} +b \left(\frac{\sigma(s)-1}{\beta}\right)^{\tfrac{n+1}{n}}$ is already taken into account. When $n=1$, we recover the Starobinsky's inflaton potential $V_E(s) = \frac{1}{4\beta}(1-e^{-\sqrt{\frac{2}{3}}s})^2$.

Indeed, the case $n=1$ is special: when we consider the large field limit, $s\gg 1$,  $\sigma(s) \gg 1$, 
\begin{align}
\lim_{s\to \infty} V_E =\frac{n}{2(n+1)\beta^{1/n}} e^{\sqrt{\tfrac{2}{3}} \left(\tfrac{1-n}{n}\right)s},
\end{align}
which approaches constant if $n=1$ so that we can realize a large field inflation scenario as Starobinsky pointed out~\cite{Starobinsky:1980te}. 

By expanding the potential the naïve cutoff scale of the theory near $s\sim 0$ becomes:
\begin{align}
V_E 
&=\alpha_n \sum_{k=1,\ell=0} \frac{(-2)^\ell (2/3)^{(k+\ell)/2}}{k!\ell!} s^{k+\ell} \\
&\equiv  \sum_{k=1,\ell=0} \frac{s^{k+\ell}}{\Lambda^{k+\ell-4}},
\end{align}
where the cutoff scales for operators with mass dimension $D=k+\ell >4$ are
\begin{align}
& \Lambda_{D} = \left[\frac{k!\ell!}{\alpha_n (-2)^\ell (2/3)^{k+\ell}}\right]^{\tfrac{1}{k+\ell-4}}
\end{align}
where $\alpha_n=\frac{n \beta^{-\tfrac{1}{n}}}{2(n+1)}$. Now requesting $\Lambda_D>1$, 
we find the lower bound on $\beta$ as
\begin{eqnarray}
\beta > \left[\frac{n 2^{2\ell+k} }{2(n+1) 3^{k+\ell} k! \ell!}\right]^n, ~~k+\ell >4.
\label{eq:unitarity}
\end{eqnarray}
As the number in the parentheses is smaller than unity in the region of our interest, the theory setup 
does not suffer from unitarity issues below the Planck scale as long as the condition in Eq.~\eqref{eq:unitarity} is satisfied.

\section{Extension of the Starobinsky model}

Given that the Starobinsky model is selected for inflation as an appropriate extension of the Einstein gravity, we introduce a cubic curvature term as the extension of the Starobinsky model, namely, take $f(R)=a R+b R^2+c R^3$. Then, we present the modified predictions for inflation in this case.~\footnote{ We note other extensions of the Starobinsky model were also studied with different perspectives~\cite{Huang:2013hsb, Sebastiani:2013eqa, Kamada:2014gma, Artymowski:2014nva}. }

Taking $a=1, b=\beta/2,$ and $c=\gamma/3$,  we get the frame function in the dual scalar theory as $f'(\phi) = 1+ \beta \phi + \gamma \phi^2$, and
\begin{eqnarray}
s(\phi) &=& \sqrt{\frac{3}{2}}\ln \left[1+ \beta \phi + \gamma \phi^2\right] , \\
\sigma(s) &\equiv&e^{\sqrt{\frac{2}{3}}s}=1+\beta \phi + \gamma \phi^2. \label{sigma}
\end{eqnarray}
The quadratic equation is easily solved and we get $\phi(s)$:
\begin{eqnarray}
\phi(s) =\frac{\beta}{2\gamma}\left(\sqrt{1+4\frac{\gamma}{\beta^2}\left(\sigma(s)-1\right)}-1 \right).
\end{eqnarray}

If $\gamma$ is small ($\gamma \ll \beta$) and $\phi\sim 1$, we may treat the $\gamma$ term as a small perturbation in $\sigma(s)$, so that we find a convenient approximation $\beta \phi(s) +1 = \sigma(s) -\frac{\gamma}{\beta^2} (\sigma(s)-1)^2+\cdots$, or
\begin{eqnarray}
\phi(s) = \frac{\sigma(s)-1}{\beta}\left[1-\frac{\gamma}{\beta}\left(\frac{\sigma(s)-1}{\beta}\right) +{\cal O}\left(\frac{\gamma}{\beta}\right)^2 \right].
\end{eqnarray}
The potential in Einstein frame is
\begin{align}
V_E(s) 
&= \frac{\beta \phi(s)^2 (1+\frac{4\gamma}{3\beta}\phi(s))}{4\left(1+\beta \phi(s)(1+\frac{
\gamma}{\beta}\phi(s))\right)^2}, \\
&\approx V_0(s) \left[1 -\frac{2}{3}\frac{\gamma}{\beta}\left(\frac{\sigma(s)-1}{\beta}\right) + \cdots \right]
\end{align}
where $V_0(s) = \frac{1}{4\beta}(1-\frac{1}{\sigma})^2 = \frac{1}{4\beta}(1-e^{-\sqrt{\frac{2}{3}s}})^2$ is the potential for $\gamma=0$. As the potential is expanded by powers of $\sqrt{2/3}s$, this setup is free from unitarity issues.

\subsection{Inflation}

The slow-roll parameters are
\begin{align}
\epsilon &= \frac{1}{2} \left(\frac{V_E'}{V_E}\right)^2  = \epsilon_0 + \frac{\gamma}{\beta} \Delta \epsilon, \\
\eta &= \frac{V_E''}{V_E} = \eta_0 + \frac{\gamma}{\beta} \Delta \eta
\end{align}
where $\epsilon_0$ and $\eta_0$ are the slow roll parameters when $\gamma=0$ and the corrections are perturbatively calculated as:
\begin{align}
\epsilon_0 &= \frac{4}{3(\sigma(s) -1)^2},\\
\eta_0 &= -\frac{4(\sigma(s)-2)}{3(\sigma(s)-1)^2},\\
\Delta \epsilon & = -\frac{8\sigma(s)}{9\beta (\sigma(s)-1)} + {\cal O}\left(\frac{\gamma}{\beta}\right),\\
\Delta \eta &=-\frac{4\sigma(\sigma+3)}{9\beta(\sigma-1)}+ {\cal O}\left(\frac{\gamma}{\beta}\right).
\end{align}

The number of efoldings from the start ($s$) to the end ($s_e < s_0$) of inflation is calculated 
\begin{align}
N_e(s) 
&=  \int_{s_e}^{s} \frac{ds}{\sqrt{2 \epsilon}} \\
&=N_{e0} + \Delta N_e,
\end{align}
where $N_{e0}$ for $\gamma=0$ and the correction term $\Delta N_e$ are
\begin{align}
N_{e0}
&= \int_{s_e}^{s} \frac{ds}{\sqrt{2\epsilon_0}} =\left[\frac{3}{4} \sigma(s) -\frac{\sqrt{6}}{4}s \right]_{s_e}^{s_0}, \\
&\approx \frac{3}{4} \sigma(s), \, (s\gg s_e\sim 1~\text{assumed})
\end{align}
and
\begin{align}
\Delta N_e 
&= -\left(\frac{\gamma}{\beta}\right)\int_{s_e}^{s} \frac{\Delta \epsilon~ d|s|}{2^{3/2}\epsilon_0^{3/2}} \\
&\approx \left(\frac{\gamma}{\beta}\right) \frac{\sigma(s)^3}{12\beta}, \, (s\gg s_e\sim 1~\text{assumed})
\end{align}
We find $s_e$ requesting ${\rm Min}(\epsilon,|\eta)|)=1$ and $s_*$ requesting 60-efoldings:
\begin{align}
N_e(s_*)= \frac{3}{4}\sigma(s_*) + \left(\frac{\gamma}{\beta}\right) \frac{\sigma(s_*)^3}{12\beta}=60.
\label{eq:60efold}
\end{align}

Finally, from the COBE normalization \cite{Akrami:2018odb},
\begin{align}
\left.\frac{V_*}{\epsilon}\right|_{s_*} 
\approx \frac{3 \sigma^2(s_*)}{16\beta} + \left(\frac{\gamma}{\beta}\right)\frac{\sigma^4(s_*)}{8\beta^2}=0.027^4.
\label{eq:COBE}
\end{align}
where $V_*$ is the inflaton vacuum energy at horizon exist,
and correspondingly determine the $R^2$ coupling as
\bea
\beta\approx 2.26\times 10^9\, \Big(\frac{N}{60} \Big)^2. \label{beta}
\eea

\begin{figure}[t]
\includegraphics[width=.489\textwidth]{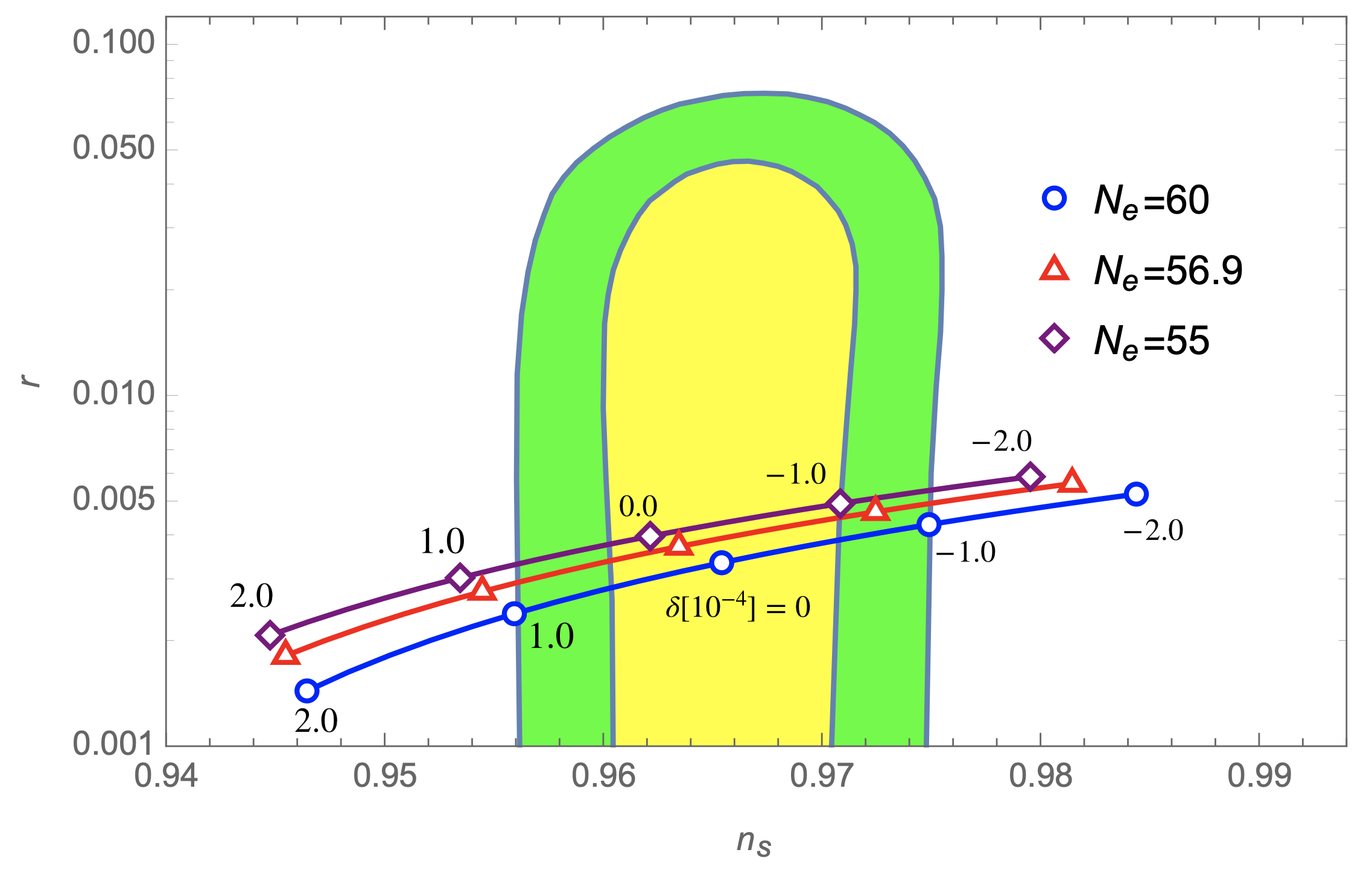}
\caption{$(n_s,r)$ for $N_e=60$(blue), $N_e=56.9$(red) and $N_e=55$(purple) efoldings with $\delta= [-2.0,2.0]\times 10^{-4}$ vs Planck2018 $1\sigma$ (Yellow) and $2\sigma$ (Green) constraints \cite{Akrami:2018odb}. 
\label{fig:ns-r}}
\end{figure}

Having two conditions from Eqs.~\eqref{eq:60efold} and \eqref{eq:COBE}, we now try to make the predictions for cosmological observations. Here we consider the spectral index and the tensor-to-scalar ratio
taking $\sigma(s_*) \approx \frac{4}{3}N_e - \delta \frac{64}{243} N_e^3$ where $\delta\equiv \gamma/\beta^2 \ll 1$ from Eq.~\eqref{eq:60efold}:
\begin{align}
n_s
&= 1-6\epsilon(s_*) + 2\eta(s_*)\\
&\approx 1-\frac{2}{N_e}-\frac{9}{2N_e^2} -\delta \frac{128}{81}N_e, \label{eq:nsapprox}
\end{align} 
and 
\begin{align}
r= 16\epsilon(s_*)
\approx \frac{12}{N_e^2} - \delta \frac{256}{27}. 
\end{align}
When $\delta=0$, we recover the well-known relations in $R^2$ inflation and consequently Higgs inflation with non-minimal coupling~\cite{Park:2008hz}
and the small $\delta$-corrections give additional contributions to observables so that we can set the bounds on the size of $\delta$ \cite{Huang:2013hsb, Pi:2017gih}.

In Fig.~\ref{fig:ns-r}, we show the effect of the $R^3$ correction with $\delta=\gamma/\beta^2\sim 10^{-4}$ in comparison with the Planck constraints in $(n_s-r)$ plane~\cite{Akrami:2018odb}. The blue, red and purple lines from bottom to top correspond for $N_e=60, 56.9$ and $55$, respectively. In particular, $N_e=56.9$, is the efolding number required for solving the horizon problem obtained by considering reheating, which we discuss in detail in the next section. Due to the negative correction to $n_s$ and the positive correction to $r$ from the positive $\delta$, the prediction moves from right-up to the left-down when $\delta$ changes from $-2.0\times 10^{-4}$ to $2.0\times 10^{-4}$. 
The middle point is for $\delta=0$ corresponding to the Starobinsky limit (or the Higgs inflation limit).

\begin{figure}[t]
\includegraphics[width=.479\textwidth]{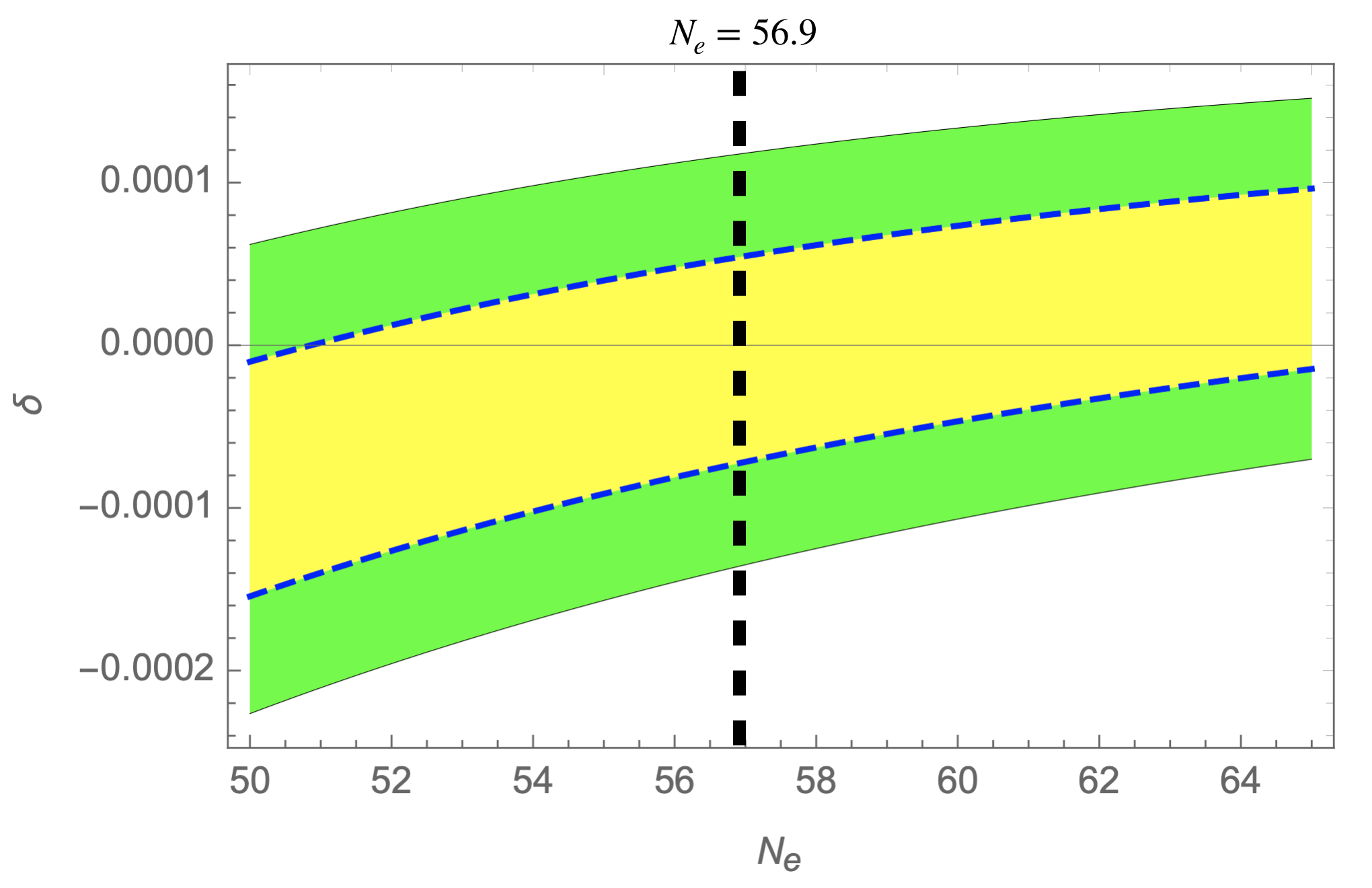}
\caption{The bound on $\delta$ for varying $N_e \in (50,65)$ from Planck2018 $1\sigma$ (Yellow) and $2\sigma$ (Green) constraints  \cite{Akrami:2018odb}. $N_e=56.9$ is indicated by the vertical, dotted line.
\label{fig:delta}}
\end{figure}

In Fig.~\ref{fig:delta}, we show the bound on $\delta$ for different choices of $N_e=50-65$  taking the Planck 2018 data into account. The vertical dotted line depicts the case $N_e=56.9$.

The running of spectral index is also calculated:
\begin{align}
\frac{dn_s}{d\log k} 
&= -2\xi +16 \epsilon \eta -24 \epsilon^2 \\
 &=-\frac{2}{N_e^2} + \frac{128}{81}\delta \left(1-\frac{153}{32 N_e^2}\right)
\end{align}
where $\xi = \frac{V_E'(s) V_E'''(s)}{V_E^2(s)}$.
The TT,TE,EE+ lowE+lensing constraint from Planck 2018 \cite{Akrami:2018odb} is 
\begin{eqnarray}
\frac{dn_s}{d\log k} = -0.0045 \pm 0.0067~(\text{$68\%$ CL}).
\end{eqnarray}
That leads $-0.0067<\left.\delta\right| _{N_e=60}<0.0017$, which gives less significant constraint at the moment. 

In passing, we comment on the initial condition for inflation in the presence of the $R^3$ corrections. In particular, for a negative value of $\gamma$, there is a potential instability developing at large inflaton field values. From eq.~(\ref{Einsteinpot}), the Einstein frame potential with the $R^3$ term included is given explicitly as a function of $\phi$ by 
\bea
V_E(\phi) = \frac{1}{4} \frac{\beta\phi^2+\frac{4}{3}\gamma\phi^3}{(1+\beta\phi+\gamma\phi^2)^2}.
\eea
The field $\phi$ is not a canonical field due to the modified kinetic term, but it is sufficient to take the above potential for the analysis of the initial condition for inflation.
Then, we find that there exists a maximum of the potential at $\phi_c=-\frac{1}{2\delta\beta}>0$ for $\beta>0$ and $\delta<0$, but it is located far beyond the regime of the slow-roll inflation near $\phi_e\sim \frac{4}{3\beta}\, N_e$, that is, $\phi_c\gg \phi_e$ for $|\delta|\sim 10^{-4}$.  Nonetheless, there might be a concern on the correct initial condition for the slow-roll inflation, $\phi_i$, because the inflaton could have rolled down to a wrong minimum for $\phi_i>\phi_c$. Therefore, we restrict ourselves to the inflaton field values satisfying $\phi_i<\phi_c$, such that the initial condition for the slow-roll inflaton is set for our previous discussion to hold.

We remark that even higher order curvature corrections such as $\frac{1}{4}\kappa R^4$ can be included, but their effects are subdominant compared to the contributions up to $R^3$ term, as far as the coefficient of the new correction term is small enough. In particular, the dual scalar theory for the extension with $R^4$ gives rise to a quartic potential, as $f(\phi)=\phi+\frac{1}{2}\beta \phi^2+ \frac{1}{3}\gamma \phi^3+\frac{1}{4}\kappa \phi^4$, thus stabilizing the scalar potential for $\kappa>0$.
For a small $\kappa$ coupling, there can be a new minimum sufficiently far away from the inflationary regime, nevertheless the inflation can roll down to a correct vacuum after inflation, being consistent with the perturbativity of the $R^4$ term.
Several studies in the literature deal with the curvature terms beyond the Starobinsky inflation model \cite{Huang:2013hsb, Sebastiani:2013eqa, Kamada:2014gma, Artymowski:2014nva} and inflation with higher curvature terms in four or higher dimensions \cite{Ellis:1998gf,Ketov:2017aau,Otero:2017thw,Aldabergenov:2018qhs}.

\section{Reheating}

In this section, we discuss the reheating dynamics in the Starobinsky model via the minimal gravitational interactions and the impact on the precise determination of the number of efoldings.

The interaction Lagrangian between the inflaton and the SM  in Einstein frame is given in terms of the trace of the energy-momentum tensor \cite{Choi:2019osi,Lee:2019aci},  as follows,
\bea
\frac{{\cal L}_{\rm int}}{\sqrt{-g}}&=&-\frac{1}{2f'(\phi)}\, T^\mu_\mu \nonumber \\
&=& -\frac{1}{2} \, e^{-\sqrt{\frac{2}{3}} s} \, T^\mu_\mu
\eea
with
\bea
T^\mu_\mu&=& -(\partial_\mu h)^2 + 4V_E + \frac{m_f}{v}\, h{\bar f} f \nonumber \\
&&-\delta_V\frac{m^2_V}{v^2}\, h^2 V_\mu V^\mu+T^\mu_{\mu,{\rm loops}}.
\eea
Here, $h$ is the Higgs boson, $f$ denotes the SM fermions, $V=W,Z$ with $\delta_V=1,2$, respectively, and $T^\mu_{\mu,{\rm loops}}$ correspond to the loop corrections due to trace anomalies \cite{Choi:2019osi}.
Expanding the inflaton near the minimum of the inflaton potential, we identify the inflaton coupling as ${\cal L}_{\rm int}= \frac{1}{\sqrt{6}}\, s\, T^\mu_\mu$. Then, assuming that electroweak symmetry is already broken at the time of reheating, the total decay rate of the inflaton with $m_s\gg m_h, m_V$ is dominated by the inflaton decay modes into the electroweak sector \cite{Choi:2019osi}, given approximately by
\bea
\Gamma_s\approx \frac{m^3_s}{48\pi M^2_P}. \label{decayrate} 
\eea
Here, from Eq.~(\ref{beta}), the inflaton mass is given by
\bea
m_s= \frac{M_P}{\sqrt{3\beta}}= 2.96\times 10^{13}\,{\rm GeV} \, \Big(\frac{60}{N_e} \Big). \label{inflatonmass}
\eea

As a result, using Eq.~(\ref{decayrate}) with Eq.~(\ref{inflatonmass}), the reheating temperature is determined from the perturbative decay of the inflaton as
\bea
T_{\rm RH} &=& \bigg(\frac{90}{\pi^2 g_*} \bigg)^{1/4} \,\sqrt{M_P \Gamma_s} \nonumber \\
&=& \bigg(\frac{100}{g_*} \bigg)^{1/4} \bigg(\frac{60}{N_e} \bigg)^{3/2} \times (4.6\times 10^9\,{\rm GeV}). \label{RH}
\eea

It is known that the number of efoldings required to solve the horizon problem depends on the reheating temperature $T_{\rm RH}$ and the equation of state $w$ during reheating \cite{Choi:2016eif}, as follows,
\bea
N_e=61.4+\frac{3w-1}{12(1+w)}\,\ln\bigg( \frac{45V_*}{\pi^2 g_*T^4_{\rm RH}}\bigg) -\ln\bigg(\frac{V^{1/4}_*}{H_*}\bigg).\nonumber \\
\eea 
In our model, the universe is dominated by matter during inflation, i.e. $w=0$. Therefore, using the results in eqs.~(\ref{RH}) and (\ref{eq:COBE}), we determine the number of efoldings as
\bea
N_e= 56.9.
\eea
Consequently, from Fig.~\ref{fig:ns-r}, we can make a definite prediction for the spectral index and the tensor-to-scalar ratio up to $R^3$ corrections.

\section{Unitarizing Higgs inflation beyond the Starobinsky model}

In this section we discuss the roles of the dual scalar field for unitarizing the Higgs inflation beyond the Starobinsky model and solving the vacuum instability problem in the SM.

 In the extended Starobinsky model with $f(R)=R+\frac{1}{2} \beta R^2 +\frac{1}{3} \gamma R^3$,
 discussed in the previous sections, we include a non-minimal coupling $\xi$ for the Higgs field $h$ in unitary gauge. Then, in the dual scalar theory, the frame function in Eq.~(\ref{frame}) becomes
\bea
\Omega^2(\phi)= 1+\beta\phi + \gamma\phi^2 +\xi h^2.
\eea
Moreover, we also add the Higgs potential in Jordan frame to get
\bea
V(\phi,h)= \frac{1}{4}\beta\phi^2 +\frac{1}{3} \gamma \phi^3+ \frac{1}{4} \lambda (h^2-v^2)^2.
\eea
Then, similarly as in Eq.~(\ref{sigma}), we make the field definition by 
\bea
\beta {\hat \sigma} = 1+\beta\phi + \gamma\phi^2 +\xi h^2.
\eea
From this, taking the $R^3$ curvature term as perturbations, the approximate solution  for $\phi$ to the above equation is given in terms of $\hat\sigma$ and $h$ by
\bea
\phi({\hat\sigma},h)= {\hat\sigma}-\frac{1}{\beta} -\frac{\xi}{\beta}\, h^2 -\frac{\gamma}{\beta} \, \Big({\hat \sigma}-\frac{1}{\beta}  -\frac{\xi}{\beta}\, h^2\Big)^2,
\eea 
in turn, leading to the Jordan frame action in a simple form,
\bea
S&=&\int d^4x \sqrt{-g} \bigg[\frac{1}{2} \beta\,\hat{\sigma} R-\frac{1}{2} (\partial_\mu h)^2 - \frac{1}{4}\beta \Big( {\hat\sigma}-\frac{1}{\beta}  -\frac{\xi}{\beta}\, h^2\Big)^2 \nonumber \\
&&\quad+\frac{1}{6}\gamma\Big( {\hat\sigma}-\frac{1}{\beta}  -\frac{\xi}{\beta}\, h^2\Big)^3- \frac{1}{4} \lambda (h^2-v^2)^2 \bigg].
\eea
This is nothing but the induced gravity model, unitarizing the Higgs inflation \cite{Giudice:2010ka,Giudice:2014toa,Ema:2017rqn,Gorbunov:2018llf}.  
By using the equation of motion for $\hat\sigma$ with ${\hat\sigma}=\frac{1}{\beta}+\frac{\xi}{\beta}\, h^2$, we can integrate out the $\hat\sigma$ field to get precisely the effective action for the Higgs inflation  \cite{Giudice:2010ka,Giudice:2014toa}. In this process, the $R^3$ curvature term maintains the same equation of motion for the $\hat\sigma$ field as in the Starobinsky model. In this regard, we can take the extended Starobinsky model as an UV completion of the Higgs inflation up to the Planck scale. As discussed in the previous sections, the robustness of the Starobinsky model for a successful inflation can be ensured in the presence of small higher curvature terms.

Finally, we remark that  the approximate potential in Einstein frame can be obtained from  $V_E=V/\Omega^4$ at the linear order in $\gamma$, as follows,
\bea
V_E&\simeq& \frac{1}{\beta^2{\hat\sigma}^2} \bigg[\frac{1}{4}\beta\Big( {\hat\sigma}-\frac{1}{\beta}  -\frac{\xi}{\beta}\, h^2\Big)^2 \nonumber \\
&&-\frac{1}{6}\gamma \Big( {\hat\sigma}-\frac{1}{\beta}  -\frac{\xi}{\beta}\, h^2\Big)^3+ \frac{1}{4} \lambda (h^2-v^2)^2\bigg].
\eea
As a result, for $\langle{\hat\sigma}\rangle\simeq \frac{1}{\beta}$,  we find that the running Higgs quartic coupling is given by
\bea
\lambda_h=\lambda+\frac{\xi^2}{\beta},
\eea
which amounts to a positive tree-level shift for $\beta>0$, ensuring the vacuum stability in the SM for a given value $\lambda$, inferred from the Higgs mass \cite{EliasMiro:2012ay,Lebedev:2012zw}, as far as the perturbativity constraint on the running Higgs quartic coupling, i.e. $\xi^2/\beta\lesssim 1$, is satisfied. 
Furthermore, the $R^3$ curvature term leads to a suppressed dimension-6 operator, ${\cal L}_{D6}=-\frac{1}{6}\, c_H h^6$ with $c_H=\gamma\,\xi^3/\beta^3=\delta\, \xi^3/\beta\lesssim \delta\, \beta^{1/2}\lesssim  4.8 (N/60)/M^2_P$ where we used $\xi^2/\beta\lesssim 1$, $|\delta|\lesssim 10^{-4}$ and Eq.~(\ref{beta}).

\section{Conclusion}

We considered an $f(R)= R +\beta R^2/2 + \gamma R^3/3$ type of gravity model for inflation. Taking the $R^3$ term as perturbations, we identified the modifications to the inflationary parameters of the original Starobinsky model. We also showed that the dual scalar theory is well defined without issues regarding unitarity below the Planck scale. The analytic expressions for the scalar spectral index ($n_s$) and the tensor-to-scalar ratio ($r$) were derived and compared with the Planck 2018 results. 
We found that the ratio of the coefficient of $R^3$ ($\gamma$) and that of $R^2$ ($\beta$) is constrained as $|\gamma/\beta^2| < 1.0 \times 10^{-4}$ at $2\sigma$ level or $0.6 \times 10^{-4}$ at $1\sigma$ level, which is consistent with the treatment of $\delta=\gamma/\beta^2$ as small perturbations in our analysis. As an important consequence of this study, we found that a slight negative $R^3$ correction to the Higgs-$R^2$ inflation may provide a better fit in $n_s-r$ plane when the primordial black hole production is significant~\cite{Cheong:2019vzl} as noticed earlier by other authors~\cite{Pi:2017gih}. 
Lastly we showed that the dual scalar field in the extended Starobinsky model is responsible for unitarizing the Higgs inflation in the presence of the non-minimal coupling for the Higgs field.

\textit{Acknowledgments}--- 
We thank Shi Pi, Misao Sasaki and Qing-Guo Huang for helpful discussions and comments. The work was initiated during the CERN-CKC Theory Institute on New Physics in Low-Energy Precision Frontier in 2020.
The work is supported in part by Basic Science Research Program through the National Research Foundation of Korea (NRF) funded by the Ministry of Education, Science and Technology (NRF-2018R1A4A1025334, NRF-2019R1A2C2003738 (HML), NRF-2019R1A2C1089334 (SCP)). 

\bibliography{refs}

\end{document}